\documentclass[ reprint,
groupedaddress,
 amsmath,amssymb,
 aps,
pra,
onecolumn]{revtex4-2}
\usepackage{graphicx}
\usepackage{epsfig}
\usepackage{epstopdf}
\usepackage{subcaption}
\usepackage{color}
\usepackage[font=footnotesize,labelformat=simple]{subcaption}

\usepackage{bm}
\usepackage{amsmath,amssymb} 
\usepackage{lineno}

\def\.{\cdot}

\def\##1{{\bf #1\mit}}
\def\_#1{{\bf #1\mit}}
\def\-#1{{\bf #1\mit}}
\def\=#1{\overline{\overline #1}}

\begin{document}
\newcommand{\red}[1]{\textcolor{red}{#1}}
\title{Temporal Interface in Dispersive Hyperbolic Media}

\author{G.~Ptitcyn$^1$}
\author{D.~M.~Sol{\'i}s$^{2,3}$}
\author{M.~S.~Mirmoosa$^4$} 
\author{N.~Engheta$^1$}
\email{engheta@seas.upenn.edu}

\affiliation{$^1$Department of Electrical and Systems Engineering, University of Pennsylvania, Philadelphia, PA 19104, U.S.A.\\
$^2$Departamento de Tecnolog{\'i}a de los Computadores y de las Comunicaciones, Universidad de Extremadura, 10003 C{\'a}ceres, Spain\\$^3$Departamento de Ingeniería Audiovisual y Comunicaciones, Universidad Polit{\'e}cnica de Madrid, 28040 Madrid, Spain\\
$^4$Department of Physics and Mathematics, University of Eastern Finland, P.O.~Box 111, FI-80101 Joensuu, Finland
}

\begin{abstract}
Spatial inhomogeneity, temporal modulation, and engineered anisotropy of parameters of electromagnetic media offer numerous opportunities for manipulating light-matter interaction over the past decades.  Here, we investigate a scenario in which we deal with the temporal interface, hyperbolic anisotropy in the form of layered structures, and frequency dispersion.  We theoretically investigate how a monochromatic uniform plane wave - propagating in an unbounded, homogeneous, isotropic dielectric medium - undergoes changes due to the rapid temporal variation of such medium into a hyperbolic dispersive medium formed by the stack of thin metal-dielectric bilayers, in which the metal follows the lossless Drude dispersion and the dielectric is assumed to be dispersionless.  We corroborate our analytical results by numerical simulations.  We observe several interesting phenomena, such as conversion of the original frequency into three pairs of frequencies, resulting in three sets of forward (FW) and backward (BW) waves.  We present the amplitudes and the time-average Poynting vectors for such FW and BW waves, and discuss some  of the salient features of such temporal interface.

\end{abstract}

\maketitle 

\section{Introduction}

Electromagnetic wave interaction with time-varying media has recently gained considerable attention and growing interest~\cite{engheta2020metamaterials, galiffi_photonics_2022, Ptitcyn2023Tutorial}.  This topic offers the notion of four-dimensional~(4D) metamaterials~\cite{engheta2023four}, wherein material parameters such as relative permittivity can undergo rapid changes in time, either independently or in conjunction with spatial variations. Interest in spatiotemporal modulation, mostly as applied to circuits, has roots dating back to the 1950s~\cite{zadeh1950determination,zadeh1950frequency,morgenthaler1958velocity,cullen1958travelling,tien1958parametric}, yet it has recently garnered significant attention in various research groups worldwide. This newfound interest is mainly due to its promising potential for unconventional wave manipulation and a diverse range of potential applications~\cite{engheta2020metamaterials,engheta2023four,lyubarov2022amplified,galiffi_photonics_2022,zhou2020broadband,pacheco2020temporal,quinones_tunable_2021,yin2022efficient,biancalana_dynamics_2007,zurita-sanchez_reflection_2009,reyes-ayona_observation_2015,lustig_topological_2018,park_spatiotemporal_2021,sharabi2021disordered,GregAtom} that include artificial magnetic field for photons~\cite{fang_realizing_2012}, optically induced negative refraction~\cite{vezzoli_optical_2018}, frequency conversion~\cite{salary2018time}, amplification~\cite{wang2018photonic,koutserimpas2018nonreciprocal}, Doppler shift~\cite{ramaccia2017doppler,ramaccia2019phase}, Fresnel drag~\cite{huidobro2019fresnel}, camouflage~\cite{liu2019time,wang2020spread}, and nonreciprocity~\cite{yu2009complete,sounas2014angular,shi2017optical,dinc2017synchronized,fleury2018non,Our2} to name a few.

A simple while fundamental scenario one may devise is a temporal interface. In analogy with a spatial interface between two different semi-infinite media, a temporal interface is defined when a spatially unbounded medium in which a wave propagates is abruptly transformed in time into another medium with different material parameters, e.g., when the relative permittivity of a medium is sharply altered in time~\cite{morgenthaler1958velocity}. Unlike conventional spatial interfaces, temporal interfaces exhibit three distinctive properties: immutability of momentum (wave vector) accompanied by a frequency shift, lack of electromagnetic energy conservation, and generation of a backward wave, that due to causality, propagates in the medium after the temporal interface~\cite{morgenthaler1958velocity,mendoncca2002time,Agrawal2014RTC}. According to these fundamental characteristics and considering complex electromagnetic media, a myriad of possibilities and opportunities for manipulation of classical or quantum fields have been uncovered, including the creation of ``wiggler mode"~\cite{jiang1975wave,wilks1988frequency,kalluri1988reflection}, temporal aiming~\cite{pacheco2020temporal}, direction-dependent wave manipulation~\cite{Sajjad2024}, inverse prism~\cite{akbarzadeh2018inverse}, antireflection temporal coatings~\cite{ramaccia2020light,pacheco2020anti,liberal2023quantum}, polarization engineering~\cite{xu2021complete,mostafa2023spin,Sajjad2024} and polarization-dependent analog computing~\cite{RizzaCastaldiGaldi23}, wave freezing and thawing~\cite{wang2023controlling,pacheco2023holding}, the transformation of surface waves into
free-space radiation~\cite{wang2023controlling,Grap19SPP}, photon-pair generation~\cite{mendoncca2000quantum,mirmoosa2023quantum}, angular-dependent inhibition of photon production~\cite{vazquez2023shaping}, photon-pair destruction and vacuum generation~\cite{mirmoosa2023quantum}. 

Despite the growing body of research in this area, the sudden creation of hyperbolic media, resulting in a temporal interface, remains largely unexplored. Here, we extend the notion of temporal interface to dispersive hyperbolic media. Specifically, we explore how a monochromatic electromagnetic uniform plane wave undergoes changes when a host medium, assumed to be a simple isotropic dispersionless dielectric (e.g., free space), is rapidly transformed into a dispersive hyperbolic medium, formed by a stack of many bilayers made of metal and dielectric layers. The frequency dispersion of hyperbolic medium is taken into account by considering the Drude dispersion for those metal layers. In particular, we reveal that a temporal interface in the presence of such anisotropy and frequency dispersion causes the splitting of the initial frequency of the wave into three pairs, which propagate mainly along the optical axes of the crystal, exhibiting canalization.

The paper is organized as follows. In Section~\ref{SECDP}, we discuss the mechanism of the corresponding temporal interface. In Sections~\ref{SECFC} and \ref{SECFA}, we explain the frequency conversion and the evolution of electromagnetic fields as the result of the temporal interface. In Section~\ref{SECNS}, we demonstrate numerical simulation results and, finally, in Section~\ref{SECCONC}, we conclude the work.



\section{Description of the Problem} 
\label{SECDP}

To start, let us consider our hyperbolic medium as an infinitely extended collection of identical bilayers, each formed by a dielectric layer of thickness $d_{\rm d}$ and relative permittivity $\epsilon_{\rm d}$ and a metallic layer of thickness $d_{\rm m}$ and relative permittivity $\epsilon_{\rm m}$.  We assumed all these bilayers are parallel with the $xz-$plane of a Cartesian coordinate system, with its $y$ axis being normal to these bilayers.  According to the effective medium theory, the elements of the relative permittivity tensor of such a medium can be written ~\cite{poddubny2013hyperbolic} as
\begin{subequations}
\begin{align}
\epsilon_{xx}&=\epsilon_{zz}=f\epsilon_{\rm m}+(1-f)\epsilon_{\rm d},\label{eq_epsxx}\\
\epsilon_{yy}&=\frac{1}{(1-f)/\epsilon_{\rm d}+f/\epsilon_{\rm m}},\label{eq_epsyy}
\end{align}
\end{subequations}
with $f=d_{\rm m}/(d_{\rm d}+d_{\rm m})$.  
Let us assume the relative permittivity of the dielectric layers not frequency dispersive, and the relative permittivity of metalic layers lossless Drude-dispersive, 
\begin{equation}
    \epsilon_{\rm m}=\epsilon_\infty-\frac{\omega_{\rm p}^2}{\omega^2},
    \label{eq_metal_permittivity}
\end{equation}
where $\omega_{\rm p}$ is the plasma frequency and $\epsilon_\infty$ is the relative permittivity at infinite frequency. Plugging Eq.~\eqref{eq_metal_permittivity} into Eqs.~\eqref{eq_epsxx} and~\eqref{eq_epsyy} and assuming $\epsilon_{\rm d}$ = $\epsilon_\infty$, we can get frequency-dependent expressions for $\epsilon_{xx}$, $\epsilon_{yy}$, and $\epsilon_{zz}$,
\begin{subequations}
\begin{align}
    \epsilon_{xx}&=\epsilon_{zz}=\epsilon_\infty-\frac{\omega_{{\rm p,eff}}^2}{\omega^2},\label{eq_effective_exx}\\
    \epsilon_{yy}&=\epsilon_\infty-\frac{\omega_{{\rm p,eff}}^2}{\omega^2-\omega_{0,yy}^2},\label{eq_effective_eyy}
\end{align}
\label{eq_effective_exx_eyy}
\end{subequations}
where
\begin{subequations}
\begin{align}
    \omega_{{\rm p,eff}}&\equiv\omega_{\rm p}\sqrt{f},\label{eq_effective_plasma_frequency}\\
    \omega_{0,yy}&\equiv\omega_{\rm p}\sqrt{\frac{1-f}{\epsilon_\infty}}.
\end{align}
\label{eq_modified_wp_w0}
\end{subequations}
In the equations above, as mentioned, it is assumed that $\epsilon_\infty=\epsilon_{\rm d}$, implying that the effective medium described in Eqs.~\eqref{eq_epsxx} and~\eqref{eq_epsyy} before the jump ($\omega_{\rm p}=0$) is an isotropic nondispersive dielectric $\epsilon_{\rm d}$, and after the jump it acquires dispersive and anisotropic properties. 
Dispersion in Eq.~\eqref{eq_effective_exx} is of the Drude type, whereas dispersion in Eq.~\eqref{eq_effective_eyy} is of the Lorentzian type, although plasma frequency is identical for both (see Eq.~\eqref{eq_effective_plasma_frequency}). Interestingly and expectedly, one can engineer effective media parameters by properly selecting the relative thicknesses of the layers.

\begin{figure}[t!]
\centering
\includegraphics[width=0.48\textwidth]{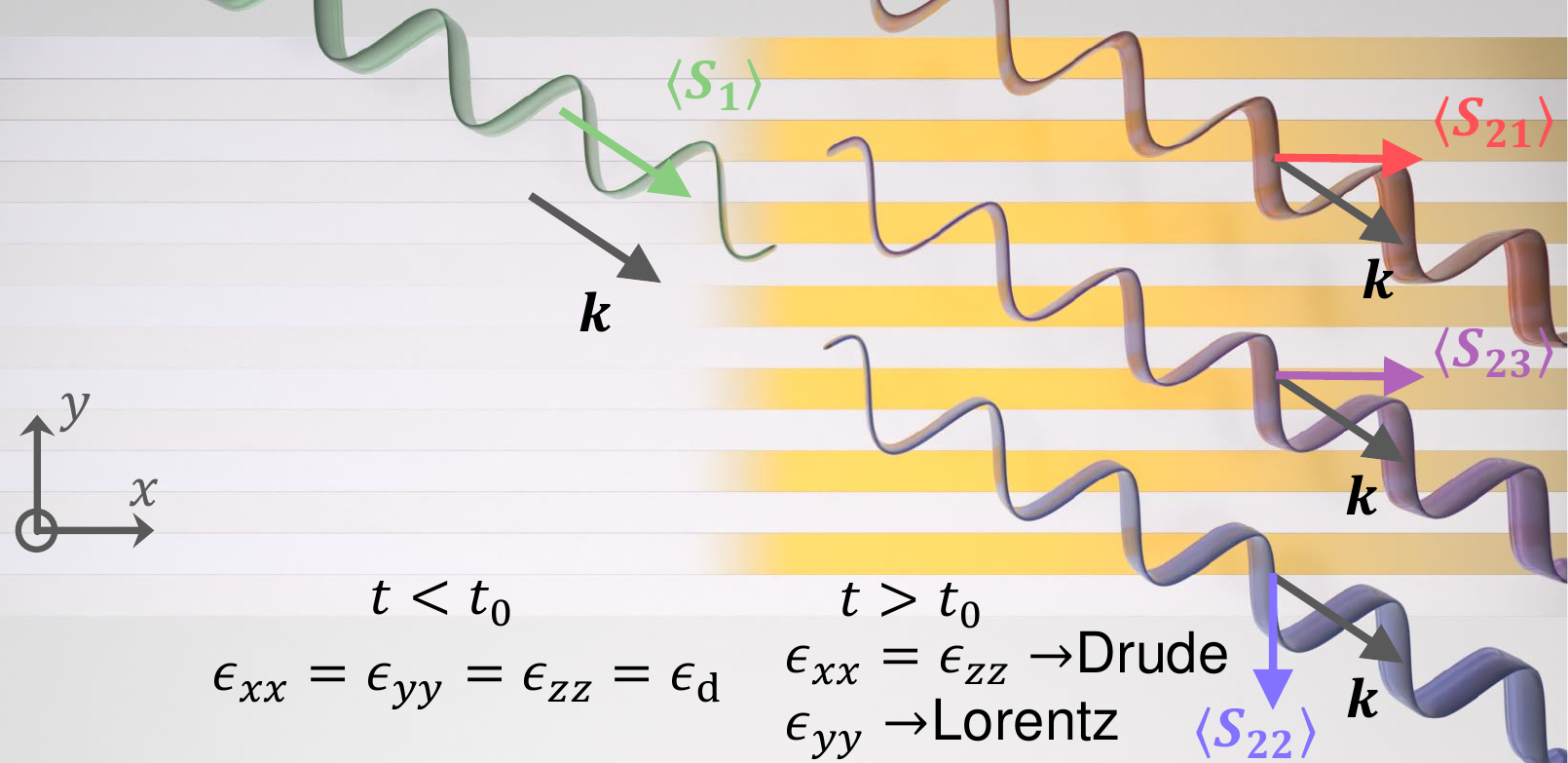}
\caption[justification=Justified]{\textbf{Temporal Interface between a simple isotopic medium and a rapidly generated dispersive anisotropic medium:}  Different colors of the waves indicate different frequencies after temporal jump. $\langle S_1\rangle$, $\langle S_{21}\rangle$, $\langle S_{22}\rangle$, $\langle S_{23}\rangle$ denote the time-average Poynting vectors of the corresponding FW waves.  We do also have time-average Poynting vector for the BW waves (not shown here for the sake of brevity)}
\label{fig_concept}
\end{figure}
%

\section{Frequency Conversion due to Temporal Interface}
\label{SECFC}

For simplicity, let us initially consider a monochromatic uniform plane wave with angular frequency $\omega_1$ propagating inside a medium described by Eq.~\eqref{eq_effective_exx_eyy} but initially parameterized with a zero plasma frequency, which makes the medium a simple isotropic and dispersionless dielectric. 
At $t=t_0$, we  abruptly ionize the metallic layers by increasing $\omega_{\rm p}$ from  zero to, say, $2\omega_1$, making the medium both anisotropic and frequency dispersive. Frequency $2\omega_1$ is chosen to show large enough contrast of the effects and, in principle, it can be chosen arbitrary. The dispersion relation for stationary (i.e., time-invariant before the time interface) anisotropic media with the electric field components in the $x-y$ plane and the magnetic field parallel with the $z$ axis reads
\begin{align}
    \frac{k_x^2}{\epsilon_{yy}}+\frac{k_y^2}{\epsilon_{xx}}=\frac{\omega^2}{c^2},
    \label{eq_disp_relation}
\end{align}
where the wave vector components $k_x$ and $k_y$ should be conserved quantities across the temporal boundary at $t=t_0$. In the absence of material dispersion, this conservation of wave vectors allows us to find, from Eq.~\eqref{eq_disp_relation}, the new converted frequency, $\omega_2$, after $t_0$~\cite{pacheco2020temporal}.
 In the case of frequency dispersive isotropic media, one obtains dispersion relations from a transcendental equation, as discussed in ~\cite{solis2021time}. In the present work, however, we have a combination of anisotropy and frequency dispersion, requiring the medium to be characterized by frequency-dependent $\epsilon_{xx}$ and $\epsilon_{yy}$, leading to another transcendental equation, which reduces to
\begin{align}    
    &k_x^2\frac{\omega_2^2-\omega_{0,yy}^2}{\epsilon_\infty(\omega_2^2-\omega_{0,yy}^2)-\omega_{{\rm p,eff}}^2}+ \nonumber\\
    &k_y^2\frac{\omega_2^2-\omega_{0,xx}^2}{\epsilon_\infty(\omega_2^2-\omega_{0,xx}^2)-\omega_{{\rm p,eff}}^2}=\frac{\omega_2^2}{c^2}.
    \label{eq_frequency_w2_1}
\end{align}
Due to anisotropy, the converted frequencies after $t_0$, $\omega_2$, depend (through $k_x$ and $k_y$) on the initial incidence angle of propagation $\psi$. Equation~\eqref{eq_frequency_w2_1} can therefore be recast as the following 6th-order characteristic equation
\begin{align} 
    &-\frac{\omega_2^2}{c^2}(\epsilon_\infty(\omega_2^2-\omega_{0,xx}^2)-\omega_{{\rm p,eff}}^2)(\epsilon_\infty(\omega_2^2-\omega_{0,yy}^2)-\omega_{{\rm p,eff}}^2)\nonumber\\
    &k_x^2\big(\epsilon_\infty(\omega_2^2-\omega_{0,xx}^2)-\omega_{{\rm p,eff}}^2\big)(\omega_2^2-\omega_{0,yy}^2)+ \nonumber\\
    &k_y^2\big(\epsilon_\infty(\omega_2^2-\omega_{0,yy}^2)-\omega_{{\rm p,eff}}^2\big)(\omega_2^2-\omega_{0,xx}^2)=0.
\end{align}
In our lossless scenario, this expression can be made 3rd order by considering $\omega_2^2$ as the new variable, with each of the three solution pairs $\pm\omega_2$ indicating a forward- and a backward- propagating wave, which hereafter will be denoted as FW and BW waves, respectively.

It is interesting to note that a temporal interface in unbounded dispersionless media (either isotropic or anisotropic) only shifts the incident frequency, resulting in a single pair of FW and BW waves with a single converted frequency (see e.g., ~\cite{morgenthaler1958velocity,pacheco2020temporal}). It has also been shown that in lossless isotropic medium with Lorentz dispersion one obtains two positive solution pairs~\cite{solis2021time}, providing two pairs of FW and BW waves with two converted frequencies. In our specific scenario here, which involves anisotropy and frequency dispersion, we observe the emergence of three solution pairs.
\begin{figure*}
\centering
\begin{subfigure}{0.24\textwidth}
 \includegraphics[width=\textwidth]{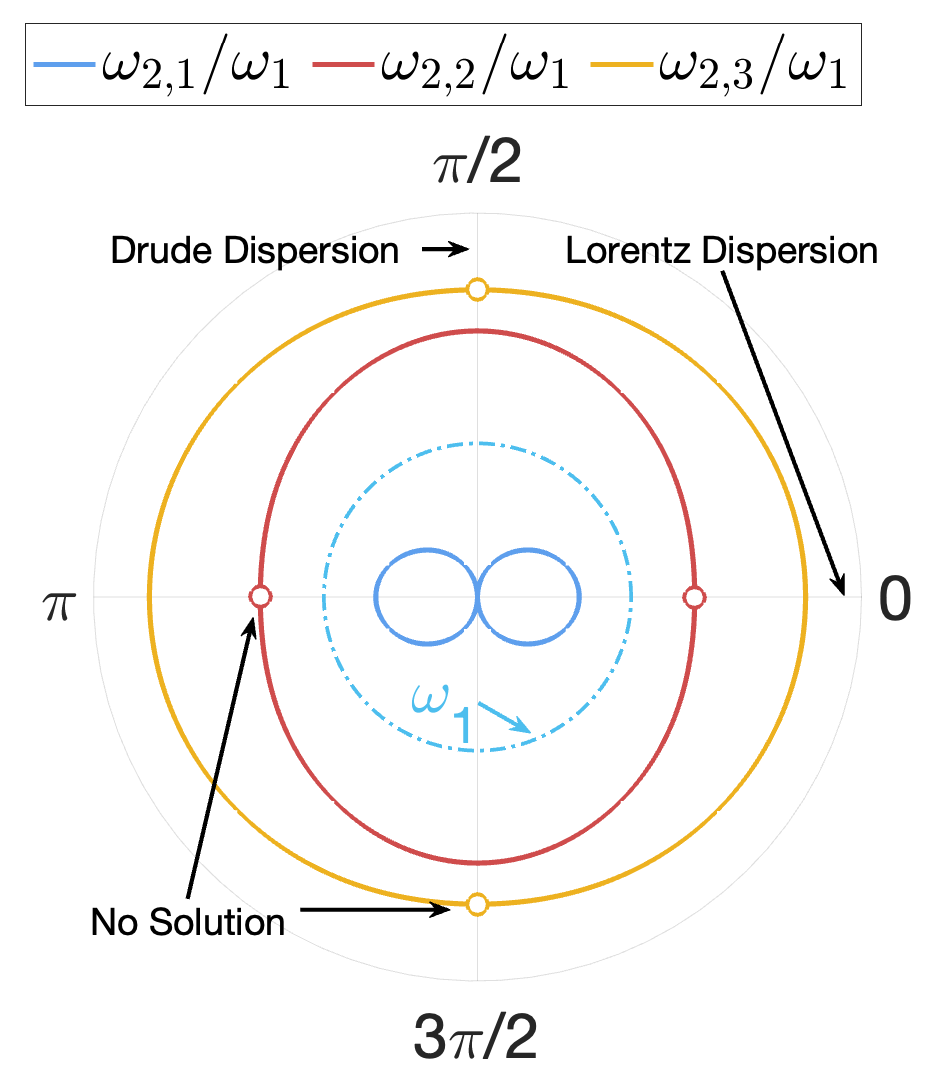}
\caption{ }
\end{subfigure}
\begin{subfigure}{0.24\textwidth}
 \includegraphics[width=\textwidth]{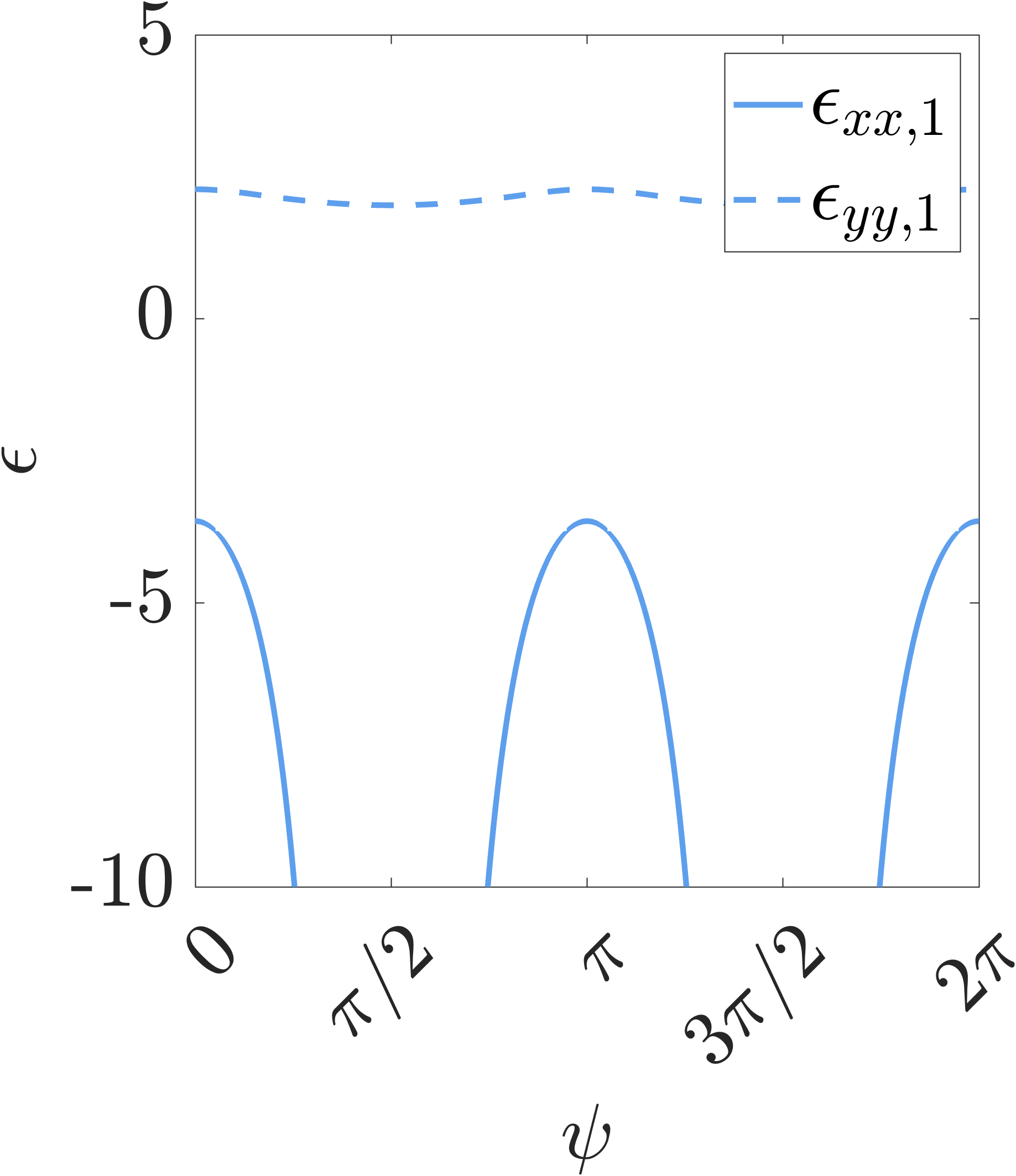}
\caption{ }
\end{subfigure}
\begin{subfigure}{0.24\textwidth}
 \includegraphics[width=\textwidth]{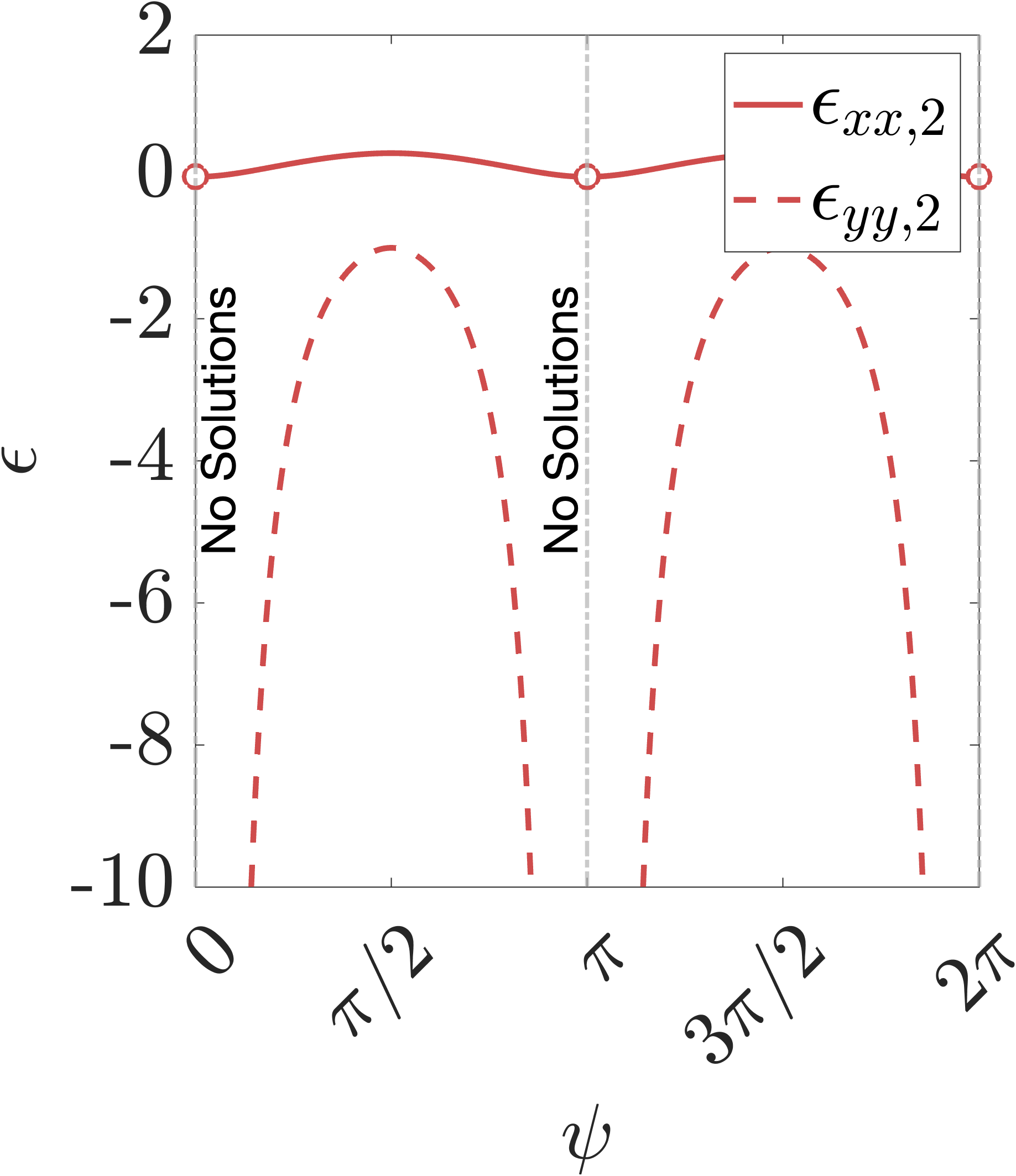}
 \caption{ }
\end{subfigure}
\begin{subfigure}{0.24\textwidth}
 \includegraphics[width=\textwidth]{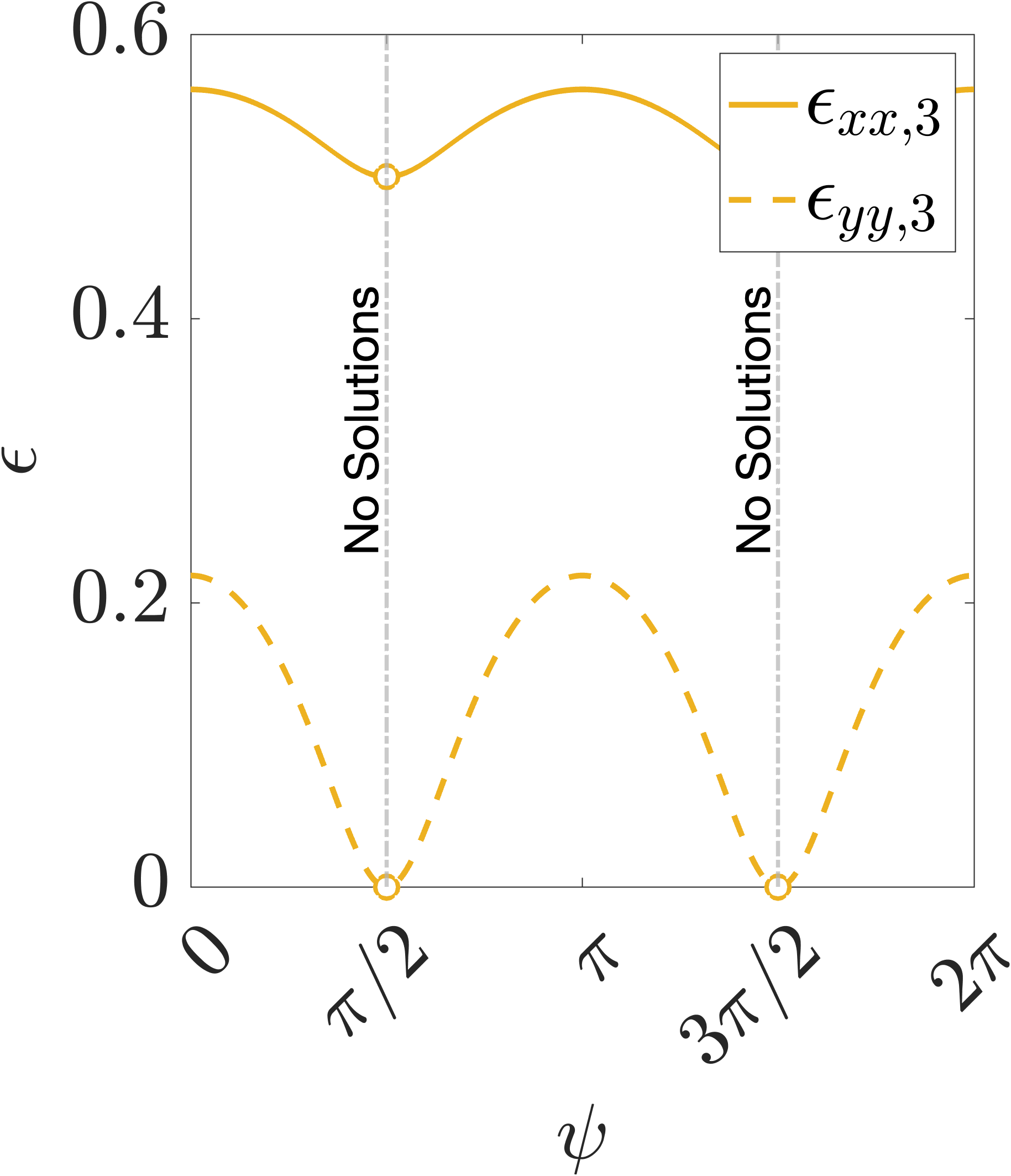}
 \caption{ }
\end{subfigure}
\caption{\textbf{Converted frequencies and relative permittivity values after temporal interface:}  (a) Polar plots of the normalized converted frequencies versus incidence angle, after sudden temporal transition from isotropic medium to anisotropic hyperbolic medium, assuming $d_{\rm d}=d_{\rm m}$, $\omega_{\rm p}=2\omega_1$ and $\epsilon_\infty=1$. (b-c) Effective relative permittivity in $x$ and $y$ directions for different converted frequencies in (a):  (b)~$\omega_{2,1}$, (c)~$\omega_{2,2}$, (d)~$\omega_{2,3}$.}
\label{fig_freq_permittivities}
\end{figure*}
Figure~\ref{fig_freq_permittivities}(a) shows the polar plots of converted frequencies after the temporal jump as a function of the direction of propagation angle of the initial wave $\psi$. Propagation along the axes is equivalent to propagation in a medium with the corresponding Lorentz or Drude dispersion. Specifically, propagation along $x$ axis ($\psi=0$ and $\psi=\pi$) or $y$ axis ($\psi=\pi/2$ and $\psi=3\pi/2$) is equivalent to propagation in a medium with permittivity $\epsilon_{yy}$ (see Eq.~\eqref{eq_effective_eyy}) or $\epsilon_{xx}$ (see Eq.~\eqref{eq_effective_exx}), respectively. Figure~\ref{fig_freq_permittivities}(a) shows three frequencies for all $\psi$ except of $\psi=m\pi/2$ with $m$ is arbitrary integer number, where the amplitude of one of the three pairs of FW and BW waves is zero.  More will be said below.


Figure~\ref{fig_freq_permittivities}(b-d) shows the $xx$ and $yy$ elements of relative permittivity tensors of the medium after the temporal interface, as evaluated for each of the three converted frequencies.  (The $\epsilon_{zz}$ is the same as $\epsilon_{xx}$.) It is evident that the medium retains its anisotropic nature; however, for each of the frequencies the nature of anisotropy is different. For $\omega_{2,1}$ and $\omega_{2,2}$, the components of the permittivity tensor exhibit opposite signs, indicating a hyperbolic nature of the medium. In contrast, at frequency $\omega_{2,3}$, all components of the permittivity tensor are positive (and less than unity), implying that the anisotropy at this frequency is of elliptic type. Additionally, at frequencies $\omega_{2,1}$ and $\omega_{2,2}$ permittivity tensors have different negative components, i.e., $\epsilon_{xx,1}$ and $\epsilon_{yy,2}$ are negative for $\omega_{2,1}$ and $\omega_{2,2}$, respectively, indicating that the isofrequency curves of the material for these two frequencies are rotated by $\pi/2$.


\section{Field Amplitudes after Temporal Interface}
\label{SECFA}

To get a better insight of the wave phenomena after the abrupt emergence of anisotropy, one needs to determine amplitudes of the fields after the temporal jump. The fields after an abrupt temporal interface consist of the three pairs of FW and BW waves. To determine their six amplitudes, we need to have six independent temporal boundary conditions.  
    \label{eq_Drude_P}
From our homogenized model of anisotropy in Eq.~\eqref{eq_effective_exx_eyy}, the second-order nature of the time-varying Drude and Lorentzain response in the form of two separate second-order differential equations for the $x$ and $y$ components of the polarization density vector $\_P$ and electric field vector $\_E$ can be written as
\begin{subequations} \label{eq_Drude_Px_Lorentz_Py}
\begin{align}
    &\frac{\mathrm{d}P_x^2(t)}{\mathrm{d}t^2}=\epsilon_0\omega_{\mathrm{p,eff}}^2(t)E_x(t), \label{eq_Drude_Px} \\
   &\frac{\mathrm{d}P_y^2(t)}{\mathrm{d}t^2}+\omega_{0,yy}^2(t)P_y(t)=\epsilon_0\omega_{{\rm p,eff}}^2(t)E_y(t), \label{eq_Lorentz_Py}
\end{align}
\end{subequations}
where $\epsilon_{\infty}$ is assumed to be unity.
Therefore, temporal continuity of $P_x$ and $P_y$ and their time derivatives at the temporal interface is implied. Together with temporal continuity of electric and magnetic flux densities, $\_D=\epsilon_0\_E+\_P$ and $\_B=\mu_0\_H$, we have a set of six temporal boundary conditions. Worth mentioning that continuity of $D_x$ and $D_y$ yields the same equations. (As an aside, it is worth noting that the temporal continuity of $\_P$ and $\_D$ automatically guarantees the temporal continuity of $\_E$).
\begin{figure*}
\begin{subfigure}{0.32\textwidth}
\centering
 \includegraphics[width=\textwidth]{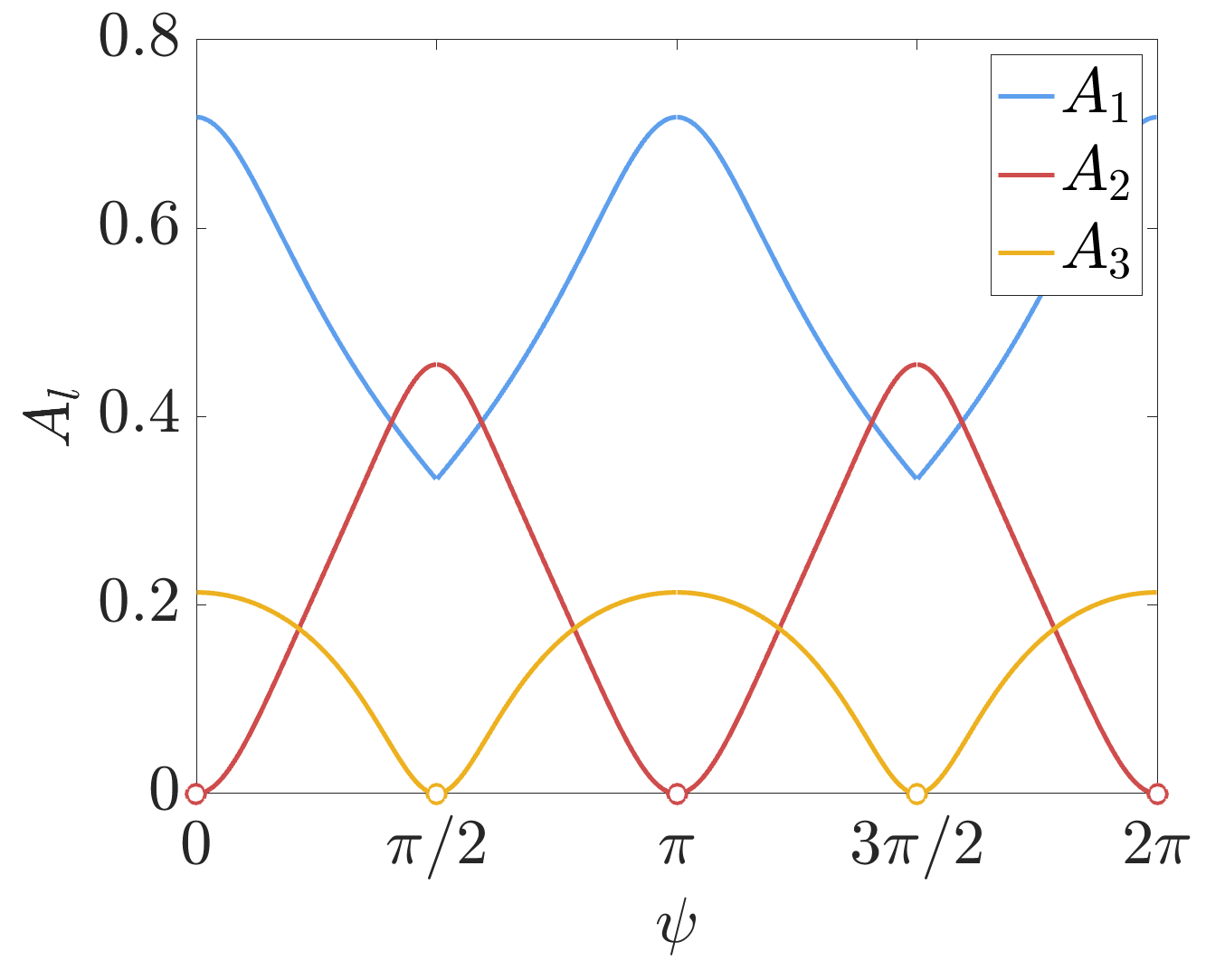}
\caption{}
\end{subfigure}
\begin{subfigure}{0.32\textwidth}
\centering
 \includegraphics[width=\textwidth]{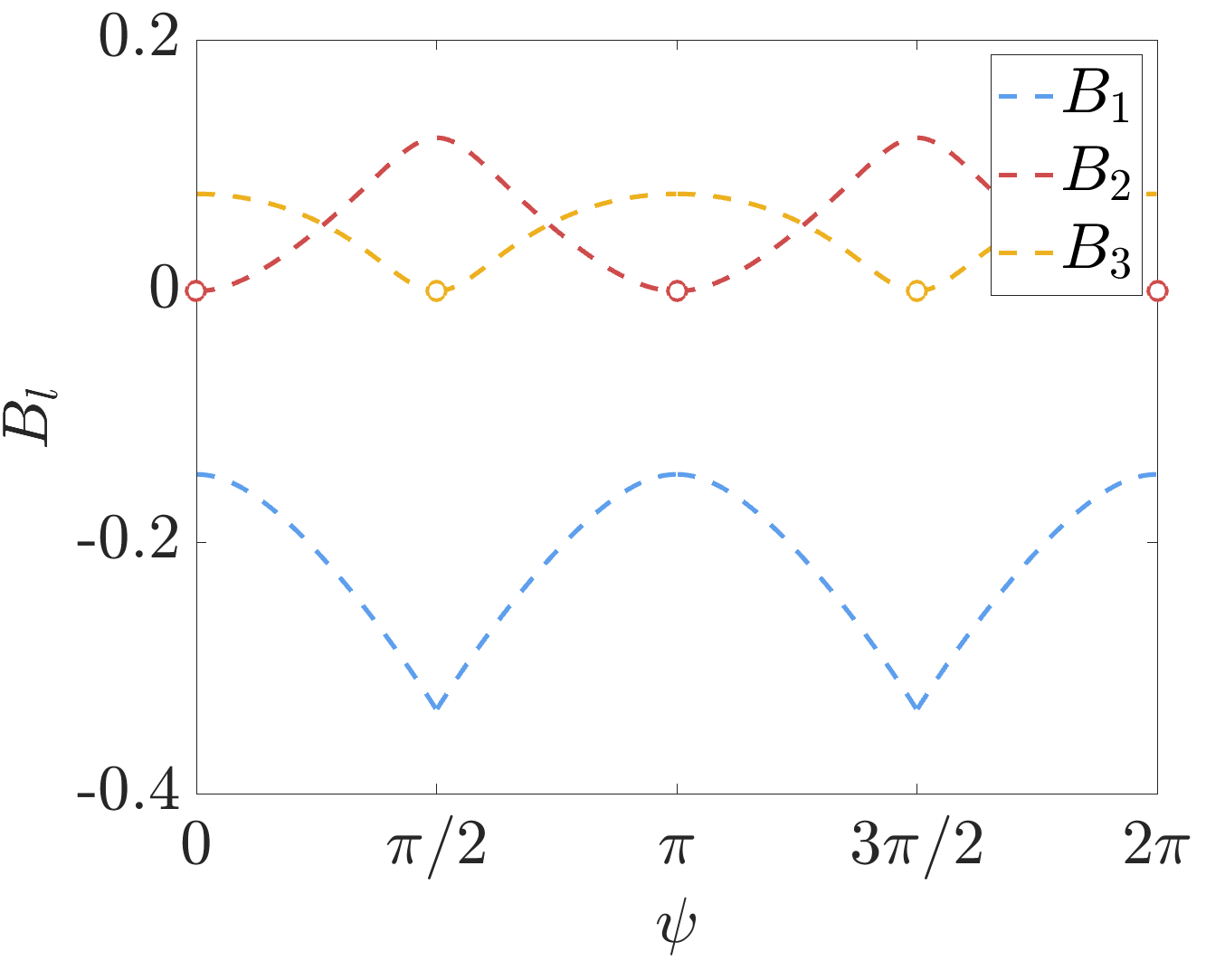}
 \caption{ }
\end{subfigure}
\\
\begin{subfigure}{0.32\textwidth}
\centering
 \includegraphics[width=\textwidth]{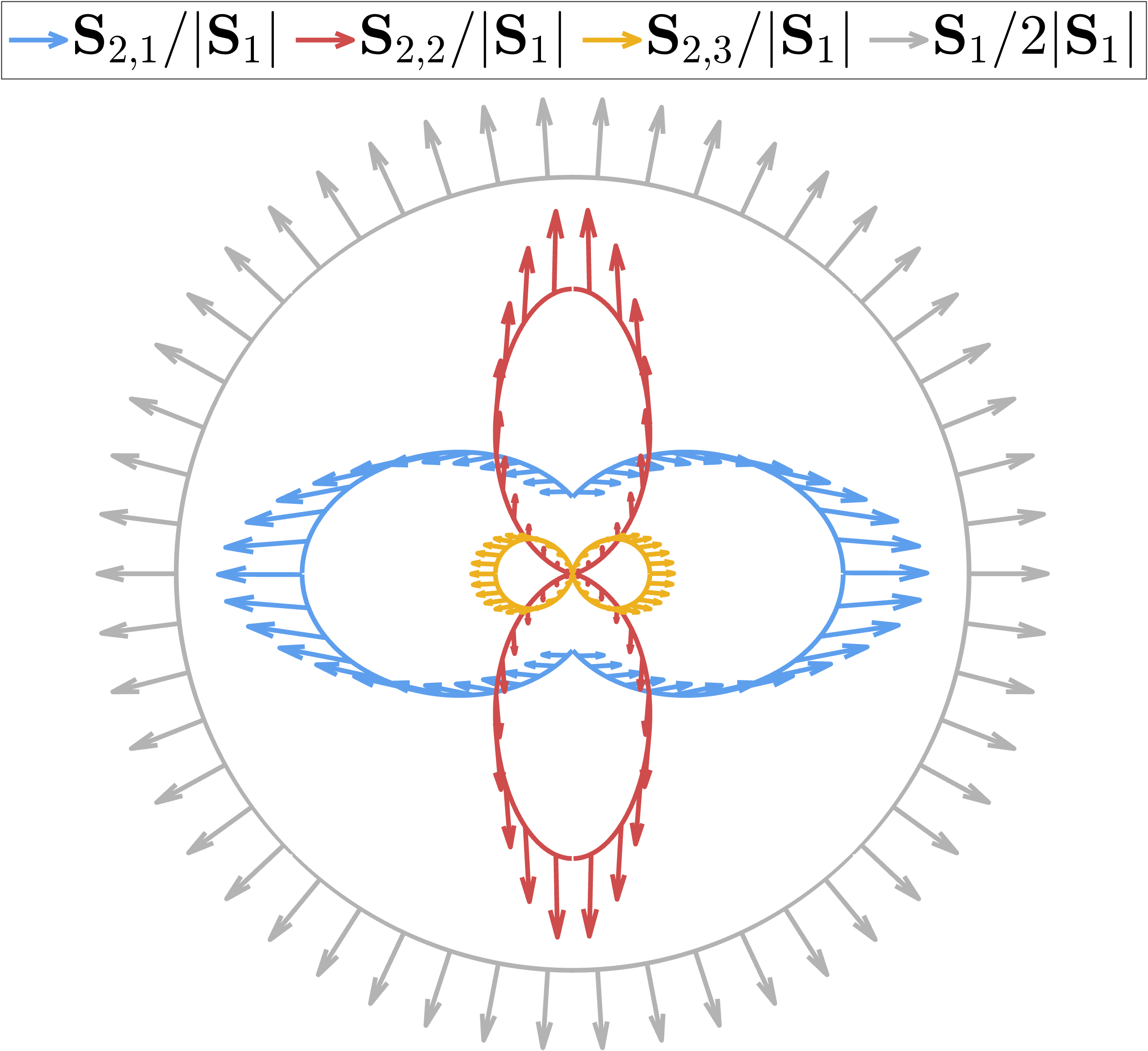}
\caption{}
\end{subfigure}
\begin{subfigure}{0.32\textwidth}
\centering
 \includegraphics[width=\textwidth]{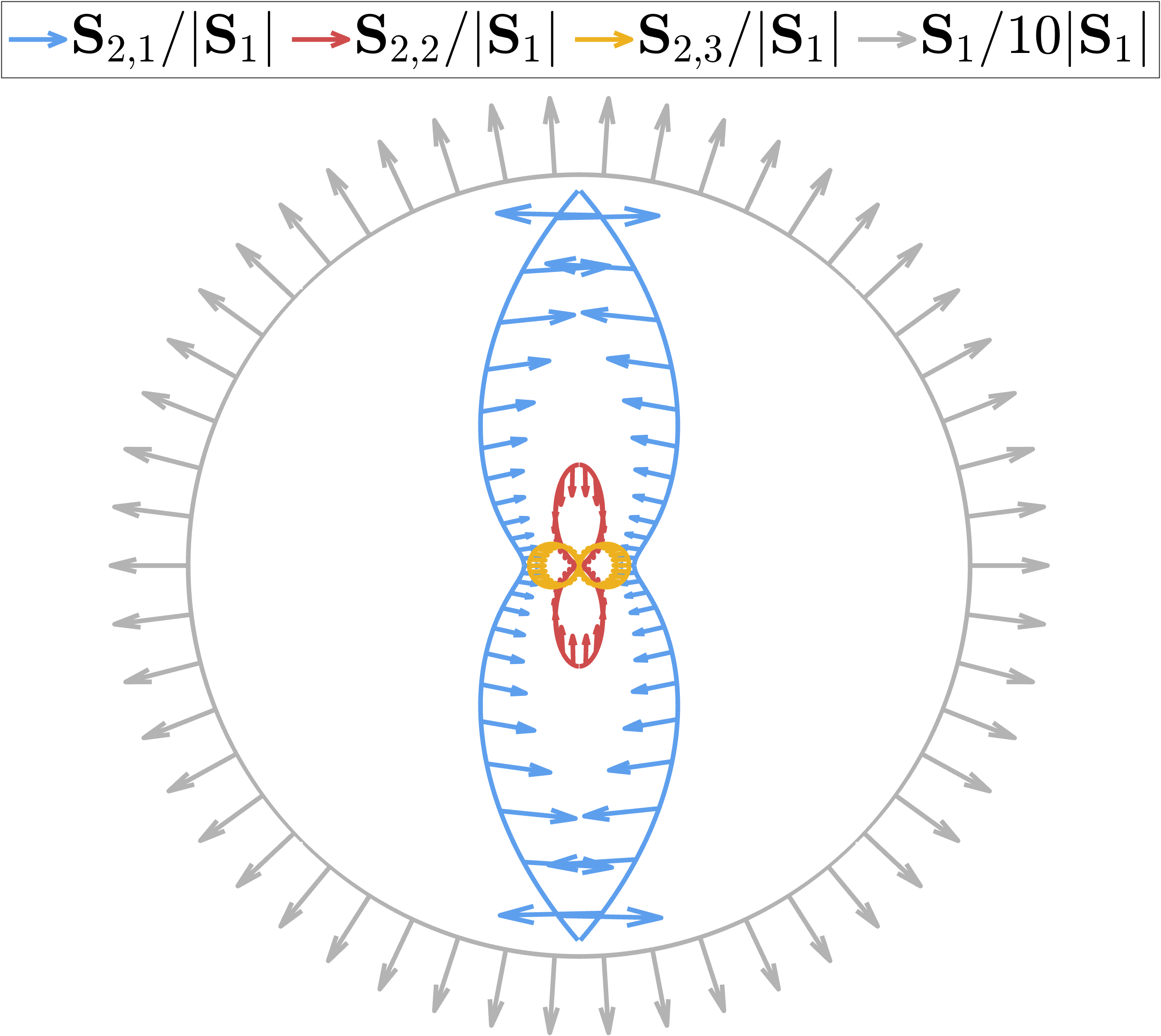}
\caption{}
\end{subfigure}
\caption{\textbf{Field amplitudes and time-average Poynting vectors after the temporal interface:}  Field amplitudes for (a) FW waves, $A_l$, and (b) BW waves, $B_l$; (c) and (d) time-average Poynting vector of FW in (c) and BW in (d), after the temporal jump, for each converted frequency as well as the initial Poynting vector (before the temporal interface) divided by 2 in (c) and divided by 10 in (d).  (Since the Poynting vectors for FW and BW waves are small compared to $S_1$, in order to show their details, we scale down $S_1$ in the figure.)}
\label{fig_scattering_coefficients_Poynting_vec}
\end{figure*}
Let us assume that before the temporal jump at $t=t_0$ we have 
\begin{subequations}\label{eq_HEDP_1}
\begin{align}
    \_H_1&=\_z e^{j\omega_1t}e^{-j\_k\cdot\_r},\label{eq_HEDP_1_H}\\
    \_E_1&=\frac{1}{\epsilon_0\omega_1\epsilon_{{\rm r},1}}(-\_x k_y +\_y k_x) e^{j\omega_1t}e^{-j\_k\cdot\_r},\label{eq_HEDP_1_E}\\
    \_D_1&=\frac{1}{\omega_1}(-\_x k_y +\_y k_x) e^{j\omega_1t}e^{-j\_k\cdot\_r},\label{eq_HEDP_1_D}\\
    \frac{{\rm d}\_P_1}{{\rm d}t}&=j\frac{\epsilon_{{\rm r},1}-1}{\epsilon_0\epsilon_{{\rm r},1}}(-\_x k_y +\_y k_x) e^{j\omega_1t}e^{-j\_k\cdot\_r},\label{eq_HEDP_1_P}
\end{align}
\end{subequations}
where subscript ``1" indicates the quantities is before $t_0$. For sake of simplicity, let us assume $t_0=0$, then quantities after $t_0$ are denoted with subscript ``2" and  they can be written as
\begin{subequations}\label{eq_HEDP_2}
\begin{align}
    \_H_2&=\_z e^{-j\_k\cdot\_r} \sum_{l=1}^{3} (A_l e^{j\omega_{2,l}t}-B_le^{-j\omega_{2,l}t}),\label{eq_HEDP_2_H}\\
    \_E_2&=\frac{e^{-j\_k\cdot\_r}}{\epsilon_0} \sum_{l=1}^{3} \bigg(-\_x\frac{k_y}{\epsilon_{xx,l}} + \_y\frac{k_x}{\epsilon_{yy,l}}\bigg) \nonumber\\
    & \frac{1}{\omega_{2,l}}(A_l e^{j\omega_{2,l}t}+B_le^{-j\omega_{2,l}t}),\label{eq_HEDP_2_E}\\
    \_D_2&= e^{-j\_k\cdot\_r}\sum_{l=1}^{3}(-\_xk_y + \_yk_x)\nonumber\\
    &  \frac{1}{\omega_{2,l}}(A_l e^{j\omega_{2,l}t}+B_le^{-j\omega_{2,l}t}),\label{eq_HEDP_2_D}\\
    \frac{{\rm d}\_P_2}{{\rm d}t}&=ie^{-j\_k\cdot\_r}\sum_{l=1}^{3}\frac{1}{\epsilon_0}\bigg(-\_x\frac{k_y(\epsilon_{xx,l}-1)}{\epsilon_{xx,l}} \nonumber\\
    &+ \_y\frac{k_x(\epsilon_{yy,l}-1)}{\epsilon_{yy,l}}\bigg)(A_l e^{j\omega_{2,l}t}-B_le^{-j\omega_{2,l}t}).
    \label{eq_HEDP_2_P}
\end{align}
\label{eq_EHDP_2}
\end{subequations}
Using equations in~\eqref{eq_HEDP_1} and~\eqref{eq_HEDP_2} and the six temporal boundary conditions one can write the following expressions:
\begin{widetext}
\begin{align}
\begin{pmatrix}
1 & -1 & 1 & -1 & 1 & -1\\
\frac{1}{\omega_{2,1}} & \frac{1}{\omega_{2,1}} & \frac{1}{\omega_{2,2}} & \frac{1}{\omega_{2,2}} & \frac{1}{\omega_{2,3}} & \frac{1}{\omega_{2,3}}\\
\frac{1}{\omega_{2,1}\epsilon_{xx,1}} & \frac{1}{\omega_{2,1}\epsilon_{xx,1}} & \frac{1}{\omega_{2,2}\epsilon_{xx,2}} & \frac{1}{\omega_{2,2}\epsilon_{xx,2}} & \frac{1}{\omega_{2,3}\epsilon_{xx,3}} & \frac{1}{\omega_{2,3}\epsilon_{xx,3}}\\
\frac{1}{\omega_{2,1}\epsilon_{yy,1}} & \frac{1}{\omega_{2,1}\epsilon_{yy,1}} & \frac{1}{\omega_{2,2}\epsilon_{yy,2}} & \frac{1}{\omega_{2,2}\epsilon_{yy,2}} & \frac{1}{\omega_{2,3}\epsilon_{yy,3}} & \frac{1}{\omega_{2,3}\epsilon_{yy,3}}\\
\frac{\epsilon_{xx,1}-1}{\epsilon_{xx,1}} & -\frac{\epsilon_{xx,1}-1}{\epsilon_{xx,1}} & \frac{\epsilon_{xx,2}-1}{\epsilon_{xx,2}} & -\frac{\epsilon_{xx,2}-1}{\epsilon_{xx,2}} & \frac{\epsilon_{xx,3}-1}{\epsilon_{xx,3}} & -\frac{\epsilon_{xx,3}-1}{\epsilon_{xx,3}}\\
\frac{\epsilon_{yy,1}-1}{\epsilon_{yy,1}} & -\frac{\epsilon_{yy,1}-1}{\epsilon_{yy,1}} & \frac{\epsilon_{yy,2}-1}{\epsilon_{yy,2}} & -\frac{\epsilon_{yy,2}-1}{\epsilon_{yy,2}} & \frac{\epsilon_{yy,3}-1}{\epsilon_{yy,3}} & -\frac{\epsilon_{yy,3}-1}{\epsilon_{yy,3}}\\
\end{pmatrix}\cdot
\begin{pmatrix}
A_1\\
B_1\\
A_2\\
B_2\\
A_3\\
B_3\\
\end{pmatrix}=
\begin{pmatrix}
1\\
\frac{1}{\omega_1}\\
\frac{1}{\omega_1\epsilon_{{\rm r,1}}}\\
\frac{1}{\omega_1\epsilon_{{\rm r,1}}}\\
\frac{\epsilon_{{\rm r,1}}-1}{\epsilon_{{\rm r,1}}}\\
\frac{\epsilon_{{\rm r,1}}-1}{\epsilon_{{\rm r,1}}}\\
\end{pmatrix}.
\end{align}
\end{widetext}
Here, each line of the matrix from top to bottom corresponds to continuity of $B_z$, $D_x$ (or $D_y$, same equation), $E_x$ (or $P_x$), $E_y$ (or $P_y$), $\frac{\mathrm{d}P_x}{\mathrm{d}t}$ and $\frac{\mathrm{d}P_y}{\mathrm{d}t}$, respectively.
Solving this system of equation provides field amplitudes for FW and BW waves. Figure~\ref{fig_scattering_coefficients_Poynting_vec}(a,b) shows linear plots of field amplitudes as a function of the direction of propagation angle $\psi$. Figure~\ref{fig_scattering_coefficients_Poynting_vec}(c and d) shows the amplitude and direction of the time-average Poynting vector of FW and BW for each converted frequency. Direction of arrows indicate the direction of the Poynting vector. Colored contours and length of the arrows indicate the amplitude of the Poynting vector of the corresponding frequency branch.  

From Figs.~\ref{fig_freq_permittivities} and~\ref{fig_scattering_coefficients_Poynting_vec}, one can make several observations: First - for the limiting cases $\psi=0$ and $\psi=\pi/2$ (and obviously also for $\psi=\pi$ and $\psi=3\pi/2$) only two sets of wave have nonzero amplitudes, since $A_2=B_2=0$ at $\psi=0$ (and $\psi=\pi$) and $A_3=B_3=0$ at $\psi=\pi/2$ (and $\psi=3\pi/2$). This results in zero time-average Poynting vector $S_{2,2}$ for FW and BW waves for $\psi=0$ (and $\psi=\pi$) and $S_{2,3}$ for $\psi=\pi/2$ (and $\psi=3\pi/2$). For all other angles of propagation we always get three pairs of FW and BW waves.  Second -- the first set for $\omega_{2,1}$  is always nonzero, i.e., $A_1\neq 0$ and $B_1\neq 0$ for all $\psi$. Third - for the first converted frequency branch, while $\epsilon_{xx,1}$ becomes negative (and infinitely large for for $\psi=\pi/2$ (and $\psi=3\pi/2$)), the corresponding wave always exists and its converted frequency $\omega_{2,1}$ approaches zero for $\psi=\pi/2$ (and $\psi=3\pi/2$), which essentially means DC field. However, we note that while the field amplitudes $A_1$ and $B_1$ at angle $\psi=\pi/2$ (and $\psi=3\pi/2$) are not zero, the $x$ component of the electric field is zero since the expression $\omega_{2,1}$$\epsilon_{xx,1}$ in the denominator approaches infinitely large value as $\omega_{2,1}$ approaching zero at $\psi=\pi/2$ (and $\psi=3\pi/2$) (see Eqs.~\eqref{eq_HEDP_2_H} and~\eqref{eq_HEDP_2_E}). Effectively for $\psi=\pi/2$ (and $\psi=3\pi/2$) we have a Drude medium, where the first solution is a DC (effectively "frozen") magnetic field (while the $xx$ element of relative permittivity for this set is infinitely negative, thus behaving as a perfect electric conductor causing the x component of the electric field to be zero) and the second solution is a propagating wave.  Fourth, it is worth noting that for waves propagating along $\psi=\pi/2$ (and $\psi=3\pi/2$) the medium after the temporal jump behaves as a Drude medium (with $A_1\neq 0$, $B_1\neq 0$, $A_2\neq 0$ and $B_2\neq 0$), but for waves propagating along $\psi=0$ (and $\psi=\pi$) the medium behaves as a Lorentzian medium (with $A_1\neq 0$, $B_1\neq 0$, $A_3\neq 0$ and $B_3\neq 0$).  For all other angles, after the temporal jump, the medium behaves as a medium with a dispersion not resembling solely Drude or Lorentzian type, with all three pairs of amplitudes $A_1\neq 0$, $B_1\neq 0$, $A_2\neq 0$, $B_3\neq 0$, $A_3\neq 0$, and $B_3\neq 0$). 
Finally, the Poynting vector in Fig.~\ref{fig_scattering_coefficients_Poynting_vec}(c and d) demonstrates another interesting insight on the wave properties. One notices that for all three converted frequencies the corresponding energy flows mainly along the optical axes, which can be attributed to a certain level of canalization of energy in this system.  We point out that along the directions $\psi=\pi/2$ and $\psi=3\pi/2$ the Poynting vector for the converted frequency $\omega_{2,1}$ (which is near zero)  is identically zero (since the $x$ component of the electric field is zero for this set). Therefore, the blue-color plots of the Poynting vector in Fig.~\ref{fig_scattering_coefficients_Poynting_vec}(c and d) have a zero value along those directions, but attain non-zero values when the direction of propagation deviates from $\psi=\pi/2$ and $\psi=3\pi/2$ with its flow being primarily along the $x$ axis.  


%




\section{Numerical Simulation} 
\label{SECNS}

\begin{figure*}
\centering
\begin{subfigure}{0.45\textwidth}
    \includegraphics[width=1\textwidth]{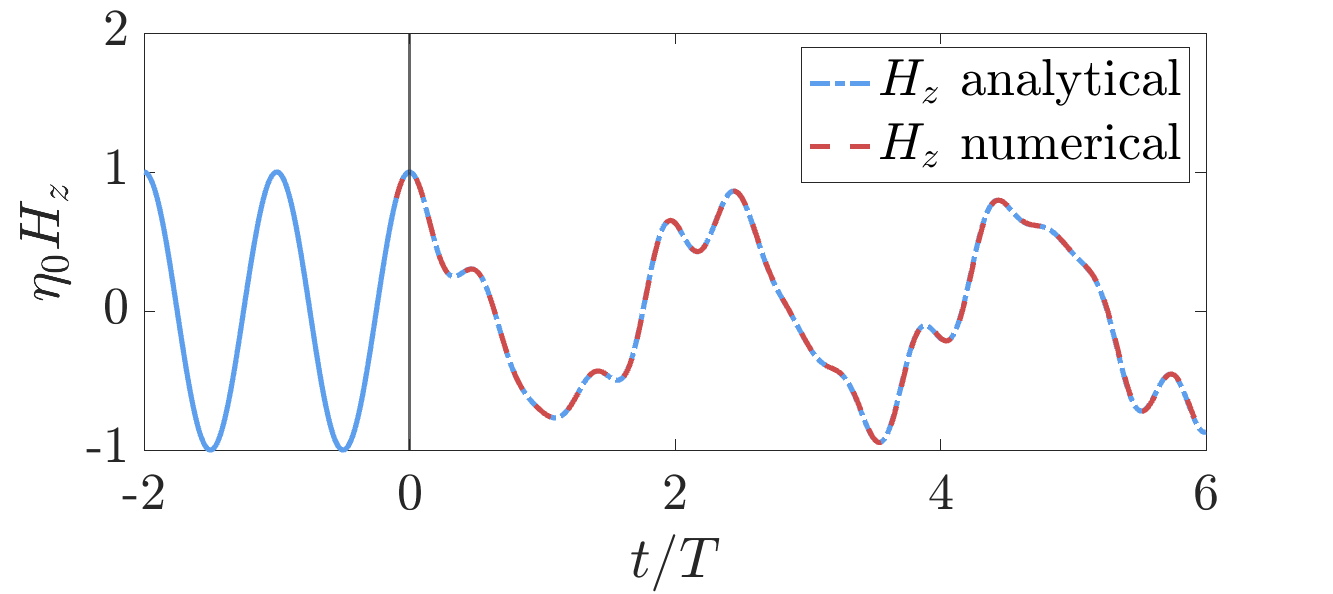}
    \caption{}
\end{subfigure}
\begin{subfigure}{0.45\textwidth}
    \includegraphics[width=1\textwidth]{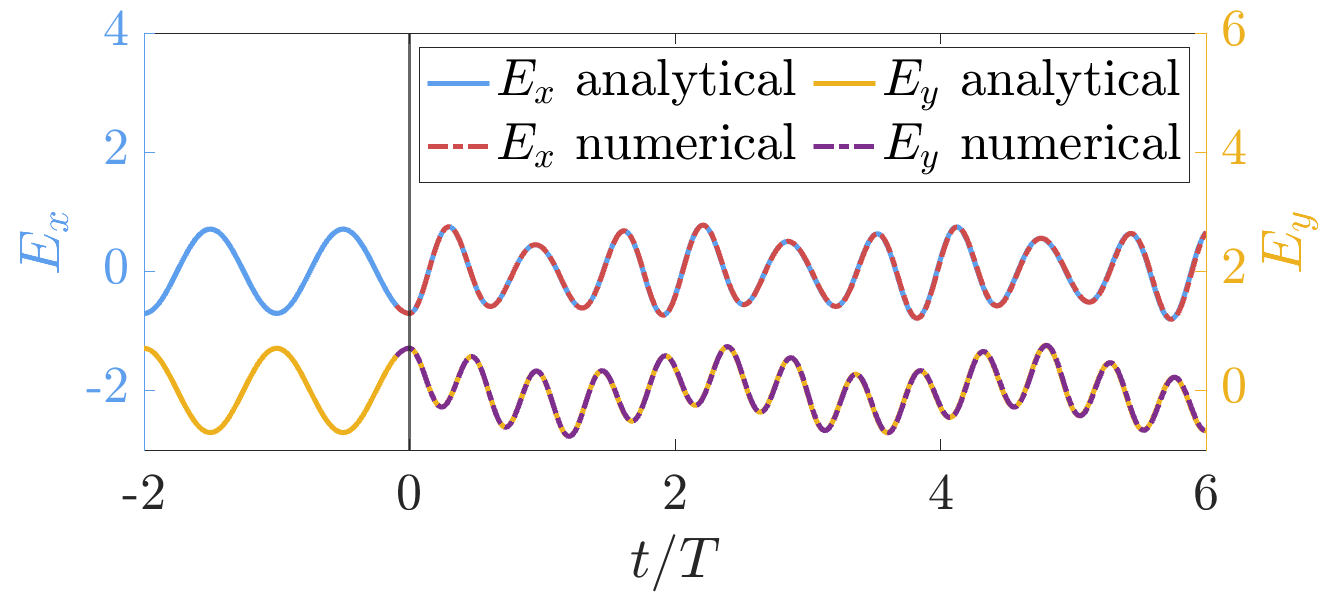}
    \caption{}
\end{subfigure}\\
\begin{subfigure}{0.45\textwidth}
    \includegraphics[width=1\textwidth]{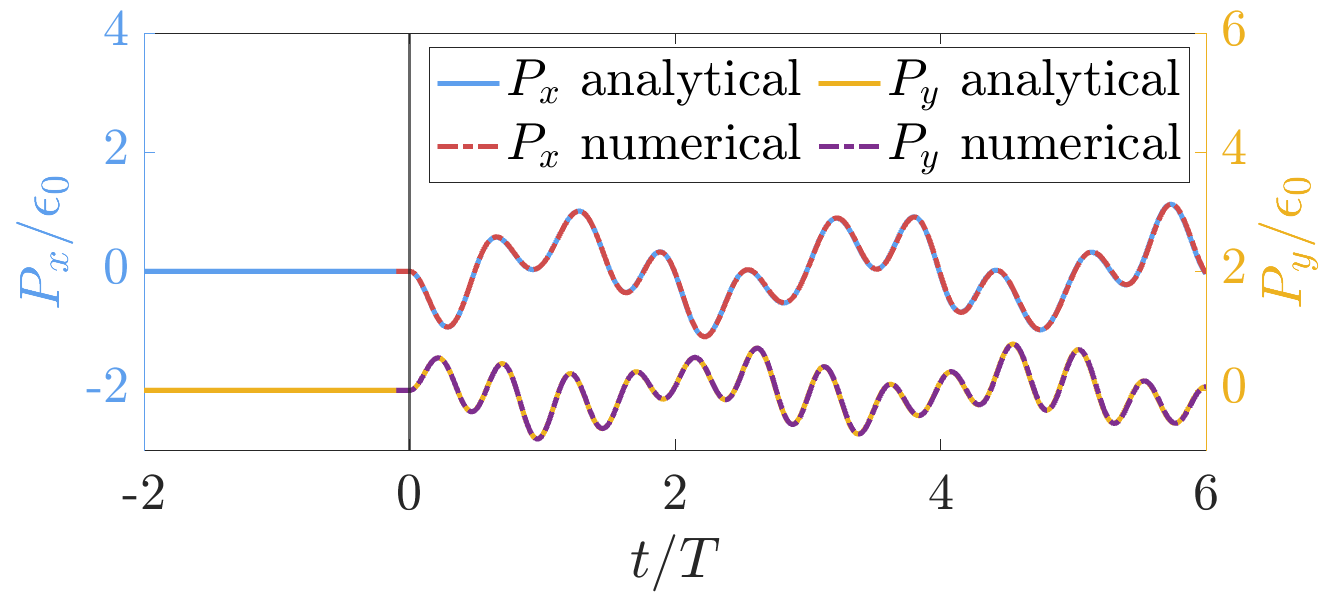}
    \caption{}
\end{subfigure}
\begin{subfigure}{0.45\textwidth}
    \includegraphics[width=1\textwidth]{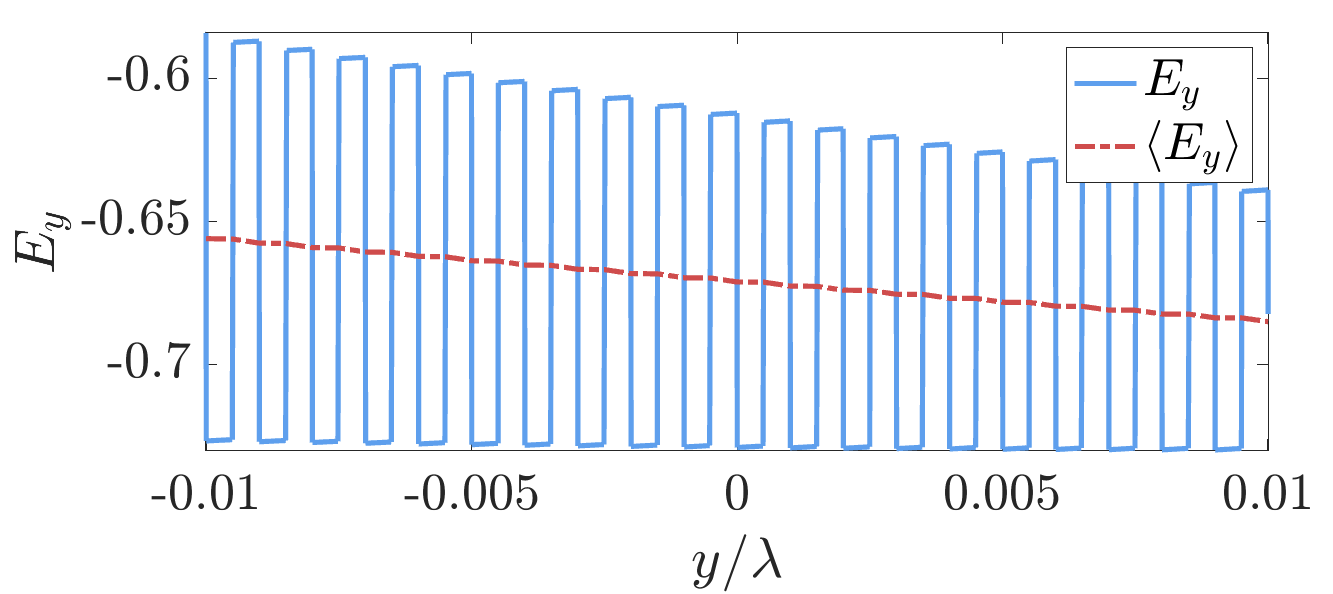}
    \caption{}
\end{subfigure}
\caption{\textbf{Comparison of analytical and numerical results}. (a-c) Show magnetic field ($H_z$), $x$ and $y$ components of electric field ($E_x$ and $E_y$) and $x$ and $y$ components for polarization density ($P_x$ and $P_y$) obtained analytically and numerically. (d) Electric field $E_y$ along $y$ axis when $x=0$ ($e^{-jk_x x}$ dependence, as seen in Eq.~(\ref{eq:A1})), which exhibits discontinuities, and its homogenized version $\langle E_y\rangle$ (resulting from $y$-averaging over a distance of $\lambda/100$ or, equivalently, about 10 unit cells) agreeing with effective medium theory.}
\label{fig_simulations}
\end{figure*}

One may argue that, rigorously speaking, the continuity of the \textit{homogenized} $\mathrm{d}\mathbf{P}/\mathrm{d}t$ adopted above might still be, at least in principle, open to debate for the following reasons: (i) it is true that, within each metal layer with the Drude dispersion, a temporal discontinuity in the plasma frequency leads to temporally continuous non-homogenized $\mathrm{d}\mathbf{P}/\mathrm{d}t$ when the Drude current response follows; and (ii) it is also clear that Eqs.~(\ref{eq_Drude_Px_Lorentz_Py}) correctly models the homogenized anisotropic response once $\omega_{\rm p}$ (and thus $\omega_{\mathrm{p,eff}}$ and $\omega_{0,yy}$) is time-invariant. But the assumption that these same equations are still indeed valid across the time interface, when $\omega_\mathrm{p}(t)$ varies in time, should be tested. To the best of our knowledge, a theory of time-varying homogenization in the presence of dispersion has not been done yet. If Eqs.~(\ref{eq_Drude_Px_Lorentz_Py}) were not to be \textit{exact}, different temporal boundary conditions might be applicable for the homogenized quantities. 
In light of this, in order to validate our analytic derivations we numerically solve homogenized model described by Eqs.~\eqref{eq_Drude_Px_Lorentz_Py} and the actual non-homogenized problem with deeply-subwavelength Drude layers. We present the results only for the latter case, since the results for the former case of homogenized model are identical with the analytical results, if the layers are thin enough ($\sim\lambda/10^3$ in our case). Regardless of the model, the transverse momentum $k_x$ allows for a dimensionality reduction, so only $y$ needs to be parameterized. We can thus write 
\begin{equation}
H_z(x,y,t)=\tilde{H}_z(y,t)e^{-jk_x x},\label{eq:A1}
\end{equation}
with phasor $\tilde{H}_z(y,t)=e^{j\omega_1 t}e^{-jk_y y}$ before the temporal transition, in unbounded vacuum. We can thus focus on the time evolution of the $k$-phasors, which abides by the curl equations
\begin{equation}
\begin{split}
    \begin{pmatrix}
        0 & -\partial_y & -jk_x \\
        \partial_y & 0 & 0 \\
        jk_x & 0 & 0
    \end{pmatrix}
    \begin{pmatrix}
        \tilde{H}_z \\
        \tilde{E}_x \\
        \tilde{E}_y
    \end{pmatrix} 
    &=
    \begin{pmatrix}
        -\mu_0 & 0 & 0 \\
        0 & \epsilon_0 & 0 \\
        0 & 0 & \epsilon_0
    \end{pmatrix}
    \partial_t \begin{pmatrix}
        \tilde{H}_z \\
        \tilde{E}_x \\
        \tilde{E}_y
    \end{pmatrix}  \\
    &+ \begin{pmatrix}
        0 \\
        \tilde{J}_x \\
        \tilde{J}_y
    \end{pmatrix},\label{eq:A2}  
\end{split}
\end{equation}
and the non-homogenized time-varying current response
\begin{equation}
\partial_t \tilde{J}_{x/y} = \epsilon_0 \omega_{\rm p}^2(y,t)\tilde{E}_{x/y},\label{eq:A3}
\end{equation}
where $\omega_\mathrm{\rm p}^2(y,t)$ is nonzero only inside the Drude layers, according to $y$. Using finite differences in both $y$ and $t$ and a Yee grid, Eqs.~\eqref{eq:A2},\eqref{eq:A3} can be solved in leapfrog fashion \cite{taflove2005computational}. Figure~\ref{fig_simulations} shows the temporal evolution of the fields at $x=y=0$ when $\psi=\pi/4$, with perfect overlap of the analytic and numerical curves, demonstrating that our assumption of temporal continuity used above for the homogenized quantities is indeed valid. An exceedingly-small unit cell of $a=\lambda/10^3$ ($y$ step of $a/200$) is chosen in order to keep the spectral content within the long-wavelength portion of the first band in the dispersion diagram, such that the homogenization behind our analytic expressions is applicable. (Anyhow, regardless of $a$, we should point out that longitudinal momentum $k_y$ is partially conserved too, in that $\tilde{H}_z(y+a,t)=e^{-jk_y a}\tilde{H}_z(y,t)$ at all times. This allows to simulate only one unit cell with periodic boundary conditions.) Accordingly, the numerical curves for $E_y$ and $P_{x/y}$, discontinuous across the successive air/metal interfaces, follow from $y$-averaging over a few unit cells. Figure~\ref{fig_simulations}(d) illustrates these discontinuities in $E_y(t=6T)$ near $y=0$.

\section{Conclusions}
\label{SECCONC}

In summary, we have extended the notion of temporal interfaces to hyperbolic frequency dispersive media. Particularly, we studied a temporal interface between vacuum and layered structure that exhibits hyperbolic properties. We discussed the interplay between anisotropy and frequency dispersion in this system, which, together with a temporal interface, results in the splitting of the original wave into three pairs with different frequencies. We discussed medium properties at the new (converted) frequencies and noticed that it leads to energy canalization, i.e. propagation of power primarily along optical axes of the structure. Finally, we conducted full wave simulations corroborating our theoretical findings.


\begin{acknowledgments}
This work is supported in part by the Ulla Tuominen Foundation to G.P., and in part by the Simons Foundation/Collaboration on Symmetry-Driven Extreme Wave Phenomena (grant SFI-MPSEWP-00008530-04) to N.E. D.M.S. acknowledges support from the Spanish Ministry of Universities under a María Zambrano Grant.

All authors have accepted responsibility for the entire content of this manuscript, consented to its submission to the journal, reviewed all the results, and approved the final version of the manuscript. N. E. conceived the idea of the paper and supervised the project, G.~P. and M.~S.~M. performed analytical modeling of the problem, G.~P. and D.~M.~S. worked on numerical verification of the obtained results.  All authors discussed the results and worked on writing and editing the manuscript. Authors state no conflict of interest.
\end{acknowledgments}

\clearpage

\end{document}